\documentstyle[twoside,fleqn,espcrc2]{article}

\newcommand{\AmS}{{\protect\the\textfont2
  A\kern-.1667em\lower.5ex\hbox{M}\kern-.125emS}}

\title{Quark Confinement in the Deconfined Phase}

\author{K. Holland 
	and
	U.-J. Wiese\address{Center for Theoretical Physics,
        Laboratory for Nuclear Science and Department of Physics, \\
        Massachusetts Institute of Technology (MIT), 
	Cambridge, Massachusetts 02139, U.S.A.}
	\thanks{Talk presented by K.H. This work is supported in 
        part by funds provided by the U.S. Department of Energy 
        (D.O.E.) under cooperative agreement DE-FC02-94ER40818}} 

\begin{document}

\begin{abstract}
In cylindrical volumes with $C$-periodic boundary conditions in the long
direction, static quarks are confined even in the gluon plasma phase due
to the presence of interfaces separating the three distinct 
high-temperature phases. An effective ``string tension'' is computed 
analytically using a dilute gas of interfaces. At $T_c$, the 
deconfined-deconfined interfaces are completely wet by the confined phase 
and the high-temperature  ``string tension'' turns into the usual string
tension below $T_c$. Finite size formulae are derived, which allow to extract
interface and string tensions from the expectation value of a single
Polyakov loop. A cluster algorithm is built for the 3-d three-state Potts 
model and an improved estimator for the Polyakov loop is constructed, 
based on the number of clusters wrapping around the $C$-periodic 
direction of the cylinder.

\end{abstract}

\maketitle

\section{INTRODUCTION}

It has long been established that the $Z(3)$ global symmetry of the 
pure $SU(3)$ gauge theory gets spontaneously broken at high temperatures
\cite{Mcle81},
leading to three distinct bulk phases, separated by interfaces.  
In this talk, we will present an unusual quark confining 
mechanism due only to the presence of these interfaces in a cylinder,
relating the free energy of a single static quark to these domain walls.
We believe that this is also evidence that the interfaces observed in 
numerical simulations correspond to physical gluonic domain walls in 
Minkowski space-time.

\section{INTERFACE GAS}

Pure Euclidean $SU(3)$ gauge theory at finite temperature 
$T = 1/\beta$ is described by the action $S[A_\mu] = \int_0^\beta dt \int 
d^3x \ (1/2 e^2) \mbox{Tr} F_{\mu\nu} F_{\mu\nu}$ and is periodic in the
Euclidean time direction. The Polyakov loop $\Phi(\vec x)$ is constructed 
from the Euclidean time component of the gauge field $A_4(\vec x,t)$ and
under gauge transformations $g(\vec x,t)$ which are also periodic in 
Euclidean time, both the action and the Polyakov loop are unchanged. 
However, if the gauge transformations differ by a center element i.e. 
$g(\vec x,t + \beta) = g(\vec x,t) z, z \in Z(3)$, the action is 
invariant but the Polyakov loop changes into $\Phi'(\vec x) = 
\Phi(\vec x) z$. The expectation value $\langle \Phi \rangle = 
\exp(- \beta F)$ measures the free energy, $F$, of a static quark. In 
the confined phase, $F$ diverges and $\langle \Phi \rangle$ vanishes, 
while in the deconfined phase $F$ is finite and $\langle \Phi \rangle 
\neq 0$. Hence, the $Z(3)$ center symmetry is spontaneously broken at 
high temperatures.

With spontaneous symmetry breaking, the spatial boundary conditions and 
the manner of the infinite volume limit are important. We 
consider a spatial volume of
size $L_x \times L_y \times L_z$. If we apply periodic boundary 
conditions, $\langle \Phi \rangle$ vanishes even 
in the deconfined phase, due to Gauss' law in a periodic volume 
\cite{Hilf83} --- 
topologically, we cannot have a single quark in a periodic box
because its center electric flux cannot go to infinity, it must end in 
an anti-quark.

We apply $C$-periodic boundary conditions in the $z$-direction only
\cite{Kron91}. 
When a $C$-periodic field is translated by $L_z$, it is replaced by its
charge conjugate. For example, for $C$-periodic gluons, 
$A_\mu(\vec x + L_z \vec e_z,t) = A_\mu(\vec x,t)^*$, where $*$ means
complex conjugate. Physically, we can have a single quark in a 
volume, partnered with an anti-quark on the other side of the 
$C$-periodic boundary. Imposing $C$-periodicity explicitly breaks the
$Z(3)$ symmetry \cite{Wies92},
 but this symmetry breaking disappears in the infinite 
volume limit. With $C$-periodic boundary conditions, $\langle \Phi \rangle$
is always non-zero in a finite volume.

The original motivation for applying $C$-periodicity in the long direction
of a cylinder was to extract the string tension $\sigma$ from numerical
simulations measuring $\langle \Phi \rangle$ in the confined phase. A single
static quark sits in the cylinder, connected by a tube of gluons to its
anti-quark partner on the other side of the $C$-periodic boundary. Then we
have
$\langle \Phi \rangle = \Sigma_0 \exp(- \beta \sigma L_z)$ and so the free
energy of a single quark in the cylinder is
\begin{eqnarray}
F = - \frac{1}{\beta} \log \Sigma_0 + \sigma L_z. \nonumber 
\end{eqnarray}
There are several advantages to using this method in a numerical study.
Firstly, it is much easier to measure the Polyakov loop than a Wilson loop 
or some other correlator. Secondly, in a periodic
box of size $L$, the Wilson loop cannot be larger than $L/2$ in its spatial
extent, whereas this technique allows us to exploit the entire volume. 

We can also investigate $\langle \Phi \rangle$ in the deconfined phase at 
temperatures $T > T_c$, where the three distinct deconfined phases 
coexist \cite{Holl97}. They are distinguished by different 
values of $\langle \Phi \rangle$ --- one expectation
value is $\Phi^{(1)} = (\Phi_0,0)$, which is rotated by $Z(3)$ 
transformations to give $\Phi^{(2)}$ and $ \Phi^{(3)}$.
A typical configuration in a cylinder consists of several bulk phases, 
aligned along the $z$-direction, separated by deconfined-deconfined 
interfaces. The interfaces cost free energy $F$ proportional to their area
$A = L_xL_y$, such that the interface tension is given by 
$\alpha_{dd} = F/A$. The
expectation value of the Polyakov loop can be calculated from a dilute gas
of interfaces \cite{Gros93}. 
The interface expansion of the partition function can be
viewed as
\begin{eqnarray}
Z &=& \begin{picture}(80,15)
\put(0,-4){\line(1,0){80}} \put(0,11){\line(1,0){80}}
\put(0,-4){\line(0,1){15}} \put(80,-4){\line(0,1){15}} \put(35,0){$d_1$}
\end{picture} \ + \
\begin{picture}(80,15)
\put(0,-4){\line(1,0){80}} \put(0,11){\line(1,0){80}}
\put(0,-4){\line(0,1){15}} \put(80,-4){\line(0,1){15}} 
\put(40,-4){\line(0,1){15}} \put(15,0){$d_2$} \put(55,0){$d_3$} 
\end{picture} \nonumber \\ 
\ &+& \
\begin{picture}(80,15)
\put(-3,-4){\line(1,0){80}} \put(-3,11){\line(1,0){80}}
\put(-3,-4){\line(0,1){15}} \put(77,-4){\line(0,1){15}}
\put(37,-4){\line(0,1){15}} \put(12,0){$d_3$} \put(52,0){$d_2$}
\end{picture} \ + ... \nonumber
\end{eqnarray}
The first term has no interfaces and thus the whole cylinder is filled with
deconfined phase $d_1$ only. An entire volume filled with either phase $d_2$ or
$d_3$ would not satisfy the boundary conditions. The second and third terms 
have one interface separating phases $d_2$ and $d_3$. Here, $C$-periodic 
boundary conditions exclude phase $d_1$. For each configuration, we integrate
the Boltzmann weight over all possible locations of the interface, then we
sum the Boltzmann weights for all configurations with all possible numbers
of interfaces. Summing the interface expansion to all orders, we obtain
$Z = \exp(- \beta f_d A L_z + 2 \gamma \exp(- \beta \alpha_{dd} A) L_z)$.
Here, $f_d$ is the bulk deconfined free energy density and $\gamma$ is a 
factor resulting from capillary wave fluctuations of the interfaces. Note
that in three dimensions, $\gamma$ is to leading order independent of the
area $A$ \cite{Priv83}.
To calculate $\langle \Phi \rangle$, we simply include the 
Polyakov loop expectation value of each configuration with the Boltzmann
weight in the interface expansion. Summing to all orders, we obtain
$\langle \Phi \rangle = \Phi_0 \exp(- 3 \gamma \exp(- \beta \alpha_{dd} 
A) L_z)$, from which we calculate that the free energy of a single static
quark in a $C$-periodic cylinder is given by
\begin{eqnarray}
F = - \frac{1}{\beta} \log \Phi_0 + 
\frac{3 \gamma}{\beta} \exp(- \beta \alpha_{dd} A) L_z. \nonumber 
\end{eqnarray}
This result is counter intuitive. Although we are in the deconfined 
phase, the quark's free energy diverges in the limit $L_z \rightarrow 
\infty$, as long as the cross section $A$ of the cylinder remains fixed.
This is the behavior one typically associates with confinement. In fact,
$\sigma' = (3 \gamma / \beta) \exp(- \beta \alpha_{dd} A)$ plays the role
of the ``string tension'', even though there is no physical string connecting
the quark to its anti-quark partner on the other side of $C$-periodic boundary.
Because the deconfined-deconfined interfaces can lead to the
divergence of a quark's free energy and have physically observable
consequences, we believe that they are more than just Euclidean field
configurations.

Because the phase transition is of first order \cite{Gava89}, 
as we approach $T_c$, the confined phase can
coexist with the three deconfined phases, so we can also have 
confined-deconfined interfaces with an interface tension $\alpha_{cd}$.
At $T_c$, there are two possibilities \cite{Frei89,Trap92}.
If we have $\alpha_{dd} = 
2 \alpha_{cd}$, a deconfined-deconfined interface always splits into two
confined-deconfined interfaces, separated by a film of confined phase 
--- this is called complete wetting. If 
$\alpha_{dd} < 2 \alpha_{cd}$, both
deconfined-deconfined and confined-deconfined interfaces are stable 
--- this is called incomplete wetting. Numerical 
simulations indicate that the gluon
system undergoes complete wetting \cite{Kaja90,Gros93}. 
In that case, the interface expansion of
the partition function is
\begin{eqnarray}
Z&=&\begin{picture}(60,15)
\put(0,-4){\line(1,0){60}} \put(0,11){\line(1,0){60}}
\put(0,-4){\line(0,1){15}} \put(60,-4){\line(0,1){15}} \put(28,0){$c$}
\end{picture} \ + \
\begin{picture}(60,15)
\put(0,-4){\line(1,0){60}} \put(0,11){\line(1,0){60}}
\put(0,-4){\line(0,1){15}} \put(60,-4){\line(0,1){15}} \put(25,0){$d_1$}
\end{picture} \ + \ \nonumber \\
&& \hspace{-10mm} \sum_i \big\{ \ \begin{picture}(60,15)
\put(0,-4){\line(1,0){60}} \put(0,11){\line(1,0){60}}
\put(0,-4){\line(0,1){15}} \put(60,-4){\line(0,1){15}}
\put(20,-4){\line(0,1){15}} \put(40,-4){\line(0,1){15}}
\put(8,0){$c$} \put(25,0){$d_i$} \put(48,0){$c$} 
\end{picture}  \ + \
\begin{picture}(60,15)
\put(0,-4){\line(1,0){60}} \put(0,11){\line(1,0){60}}
\put(0,-4){\line(0,1){15}} \put(60,-4){\line(0,1){15}}
\put(20,-4){\line(0,1){15}} \put(40,-4){\line(0,1){15}}
\put(5,0){$d_i$} \put(28,0){$c$} \put(45,0){$d^*_i$}
\end{picture} \ \big\} \ + ...  \nonumber
\end{eqnarray}
The sum over $i$ extends over the three deconfined phases and $d^*_i$ 
denotes the charge-conjugate of $d_i$. For example, $d^*_1 = d_1$ and 
$d^*_2 = d_3$. Note that due to complete wetting, one always has an 
even number of interfaces. As before, we sum the expansion to all orders
to calculate the partition function and $\langle \Phi \rangle$, 
obtaining
\begin{eqnarray}
\langle \Phi \rangle \hspace{-1mm} = \hspace{-1mm} 
\frac{\Phi_0 \exp(\beta x L_z) + 
\Sigma_0 \exp(- \beta (x + \sigma) L_z)}
{2 \cosh[\beta L_z \sqrt{x^2 + (3 \delta^2/\beta^2) 
\exp(- 2 \beta \alpha_{cd} A)}]}. \nonumber 
\end{eqnarray}
Here, $\delta$ is the factor characterizing capillary wave fluctuations
of the confined-deconfined interfaces and $x = \frac{1}{2} (f_c - f_d) A$
measures the bulk free energy difference between confined and deconfined 
phases. From this expression, we can extract an effective ``string 
tension'' $\sigma'$ which, as we lower the temperature further into the
confined regime, reduces to the standard string tension $\sigma$ in the
large $A$ limit, as expected. 

We can also consider the incomplete wetting scenario by including 
configurations with deconfined-deconfined interfaces at $T 
\approx T_c$. The effective ``string tension'' we extract again matches
$\sigma$ as we move into the confined regime. 

\section{IMPROVED ESTIMATOR}

Using the analytically derived finite size formulae, we can determine
the interface tensions $\alpha_{dd}$ and $\alpha_{cd}$ from numerical
simulations. Because simulating $SU(3)$ gauge theory is computationally
intensive, we select a simpler model, the $3$-d three-state Potts model.
Its action is given by
$S = - \beta \sum_{x,\mu} \delta_{\Phi_x,\Phi_{x+\hat{\mu}}}$,
where $\Phi_x \in Z(3)$. This model has a $Z(3)$ symmetry and a first
order phase transition between three distinct ordered (deconfined) 
phases and one disordered (confined) phase. Here, $\langle \Phi
\rangle$ corresponds to the average spin. We have developed a single
cluster algorithm with an improved estimator for measuring $\langle \Phi
\rangle$ in a cylinder with $C$-periodicity imposed in the long 
direction. As we grow the cluster, we keep account of how many times the
cluster crosses the $C$-periodic boundary. If the cluster wraps around 
the cylinder in the long direction an 
odd number of times, we count a 1. Otherwise, we have a non-wrapping
cluster and we count a 0. This is because a non-wrapping cluster is not
aware of the $C$-periodicity and so can flip to any of the three 
deconfined phases, making on average a zero contribution to $\langle 
\Phi \rangle$, whereas a wrapping cluster can only be in  
deconfined phase $d_1$ satisfying $C$-periodicity. Because we 
have an improved
estimator, we can accurately determine very small quantities. Work on
simulations to extract the interface tensions is in progress.

\end{document}